\documentclass[aip, pop, reprint]{revtex4-1}

\usepackage[utf8]{inputenc}
\usepackage[T1]{fontenc}
\usepackage[pdftex]{graphicx}
\usepackage{amsmath}
\usepackage{amsfonts}
\usepackage{amssymb}
\usepackage{amsxtra}
\usepackage{amsthm}
\usepackage{mathrsfs}
\usepackage{color}
\usepackage{lipsum}
\usepackage{placeins}

\begin{document}

\title{Larmor radius effect on the control of chaotic transport in tokamaks} 
\author{L.A. Osorio-Quiroga}
\email{losorio@usp.br}
\affiliation{University of São Paulo, Institute of Physics, 05508-090 São Paulo, SP, Brazil}
\author{M.Roberto}
\altaffiliation{In memory of Marisa Roberto, an invaluable collaborator whose significant contributions to our work will be greatly missed.}
\affiliation{Aeronautics Institute of Technology, Physics Department, \mbox{12228-900 São José dos Campos, SP, Brazil}}
\author{R.L. Viana}
\affiliation{Federal University of Paraná, Physics Department, \mbox{81531-990 Curitiba, PR, Brazil}}
\author{Y. Elskens}
\affiliation{Aix-Marseille Université, UMR 7345 CNRS, PIIM, 13397 Marseille, France}
\author{I.L. Caldas}
\affiliation{University of São Paulo, Institute of Physics, 05508-090 São Paulo, SP, Brazil}

\date{\today}

\begin{abstract}

We investigate the influence of the finite Larmor radius on the dynamics of guiding-center test particles subjected to an $\mathbf{E} \times \mathbf{B}$ drift in a large aspect-ratio tokamak. For that, we adopt the drift-wave test particle transport model presented by W. Horton [Physics of Plasmas \textbf{5}, 3910 (1998)] and introduce a second-order gyro-averaged extension, which accounts for the finite Larmor radius effect that arises from a spatially varying electric field. Using this extended model, we numerically examine the influence of the finite Larmor radius on chaotic transport and the formation of transport barriers. For non-monotonic plasma profiles, we show that the twist condition of the dynamical system, i.e.,\ KAM theorem's non-degeneracy condition for the Hamiltonian, is violated along a special curve, which, under non-equilibrium conditions, exhibits significant resilience to destruction, thereby inhibiting chaotic transport. This curve acts as a robust barrier to transport and is usually called shearless transport barrier. While varying the amplitude of the electrostatic perturbations, we analyze bifurcation diagrams of the shearless barriers and escape rates of orbits to explore the impact of the finite Larmor radius on controlling chaotic transport. Our findings show that increasing the Larmor radius enhances the robustness of transport barriers, as larger electrostatic perturbation amplitudes are required to disrupt them. Additionally, as the Larmor radius increases, even in the absence of transport barriers, we observe a reduction in the escape rates, indicating a decrease in chaotic transport.
\end{abstract}

\maketitle 

\section{Introduction}\label{sec:Introduction}

It is well-known that the transverse transport coefficients of tokamak plasmas predicted by the neo-classical theory are much smaller than the experimental results by one order of magnitude or more. This discrepancy is commonly referred to as anomalous transport\cite{wootton1990}. \textcolor{black}{Electrostatic drift turbulence, dominated by the $\mathbf{E}\times\mathbf{B}$ drift and low-frequency waves, is a plausible candidate for explaining the high levels of particle and heat loss in tokamaks\cite{manfredi1997, conner1994}}.\par 

\textcolor{black}{Controlling transport in magnetically confined plasmas is crucial for advancing toward the goal of achieving controlled thermonuclear fusion. One area of particular interest is the study of impurity transport. Understanding the transport mechanisms of these particles is essential since impurities are unavoidable and can significantly impact plasma performance\cite{angioni2021}. Specifically, impurity accumulation in the plasma core can lead to cooling of the hot core through radiation loss. But, in the divertor, the accumulation can be advantageous, as it helps to distribute heat over a larger area, thus reducing potential damage to the wall\cite{jensen1977,shimada1991, gravier2019}.} 

\textcolor{black}{In particular, turbulence caused by drift waves plays a major role in driving the impurity flux\cite{futatani2010}.  Additionally, because the Larmor radii of impurities can be much larger than those of thermal ions \cite{kim1994}, the response of impurities to drift turbulence is expected to differ \cite{manfredi1997}. In this regard, theoretical estimations suggest that the quasilinear impurity flux is reduced for heavier particles\cite{futatani2010}, i.e.,\ large Larmor radii. However, this formulation holds only when overlapping resonances occur in the Hamiltonian motion of test particles\cite{elskens2003}.}

\textcolor{black}{In this context, test particle approaches, i.e.,\ valid when impurities are sufficiently diluted so as not to affect the turbulence, have proven useful in studying key transport mechanisms\cite{marcus2008, toufen2012}. In particular, passive tracers of impurities driven by the $\mathbf{E}\times\mathbf{B}$ drift in 2D electrostatic drift turbulence exhibit a reduction in transport levels as the Larmor radius increases. This reduction occurs because the large Larmor radius effect averages out the smaller-scale components of the electrostatic field, effectively suppressing their influence\cite{manfredi1996, dewhurst2010}. Thus, considering the finite Larmor radius (FLR) effect is essential for correctly estimating the transport properties of particles\cite{martinell2013, fonseca2014}.}\par

\textcolor{black}{The effect of the FLR on chaotic transport has been studied using discrete gyro-averaged area-preserving maps. This is possible since particle advection in a turbulent electrostatic field with a strong magnetic field can be associated with Hamiltonian dynamical systems, based on the guiding-center motion approximation due to the $\mathbf{E}\times\mathbf{B}$ drift velocity\cite{Chandre2010, marcus2008}. These models are particularly valuable as they enable the integration of particle orbits over long transport timescales. In particular, it has been shown that the probability of a particle remaining trapped in a drift-wave resonance increases when the FLR increases, improving the particle confinement \cite{fonseca2014, fonseca2016}. The inclusion of the FLR effect changes the properties of transport since it leads to chaos suppression\cite{manfredi1996, fonseca2014, kryukov2018}.}\par   

\textcolor{black}{In the presence of internal transport barriers (ITBs), created by a reversed magnetic shear configuration or external $\mathbf{E}\times\mathbf{B}$ shear flow\cite{ida2018}, the movement of impurities toward the plasma core is blocked, reducing the inward turbulent transport of impurities that are produced on the wall. Furthermore, ITBs are favorable configurations, as they are associated with some mechanisms of decontamination of the plasma core\cite{futatani2010b}. This is particularly significant, as the formation of ITBs not only limits impurity transport but also strengthens plasma confinement, helping to prevent degradation\cite{garbet2005}.}

In this work, we \textcolor{black}{examine the effect that the FLR has} on a specific type of internal transport barrier known as the shearless transport barrier (STB). This barrier can arise in non-monotonic plasma profiles configurations, such as the safety factor, the radial electric field, or the toroidal plasma velocity\cite{caldas2012}. These configurations can lead to non-twist behavior, for which the twist condition of the dynamical system, \textcolor{black}{i.e.,\ KAM theorem's non-degeneracy condition for the Hamiltonian}, is violated on a special curve (the STB), where the angular frequency of motion reaches an extremum\cite{morrison2000}. \textcolor{black}{Specifically, the onset of STBs has been proposed as a plausible mechanism for transport reduction in both the Tokamak Chauffage Alfv\'en Br\'esilien (TCABR)\cite{marcus2008, osorio2023edge} and the Texas Helimak\cite{toufen2012}, in discharges where a biased electrode at the plasma edge induces a reversed shear configuration of the electric field.} \par 

From a dynamical point of view, non-twist systems have an unusual behavior because standard results, such as KAM theory, Chirikov stochasticity criterion, etc. may not be valid\cite{delCastillo1996}. The degeneracy of the system allows the formation of twin (dimerized) islands which, as a non-linearity parameter is varied, do not overlap and break down as they usually do in twist systems. Instead, twin islands experience a kind of reconnection associated with the existence of a shearless curve that prevents the formation of a large chaotic region\cite{wurm2005, shinohara1998}. The shearless curve acts as a robust barrier to transport since it is resilient under variations of the non-linearity parameter; only strong perturbations can disrupt it\cite{szezech2009, szezech2012}.\par

\textcolor{black}{Based on this, several studies have examined the effect of the FLR on chaotic transport using a non-twist dynamical system description. Specifically, it has been shown that as the FLR varies, STBs become more resilient to disruption and can undergo bifurcations, highlighting the FLR influence on phase space topology\cite{martinell2013, fonseca2014}. Furthermore, super-diffusive behavior in the plane perpendicular to the magnetic field is associated with the presence of a STB, which emerges as the FLR increases\cite{stanzani2023}. These studies suggest that the FLR effect inhibits chaotic transport and that increasing the FLR can lead to the restoration of STBs.}\par

\textcolor{black}{In this work, we adapt the drift-wave guiding-center test particle transport model from Ref.~\onlinecite{horton1998} by introducing a suitable second-order gyro-averaged extension, which accounts for the finite Larmor radius effect that arises from a spatially varying electric field.} In particular, the model from Ref.~\onlinecite{horton1998} without this extension has been explored in recent studies addressing different features of chaotic transport of impurities when STBs are present\cite{souza2024,grime2023, osorio2023ExB}. In most of these works, we assume that the Larmor radius is too small to play a significant role in the dynamics. However, this assumption is no longer valid when considering fast particles, e.g.,\ alpha particles, which tend to accumulate in the core of burning plasmas\cite{angioni2021}.\par  

\textcolor{black}{Hence, with the extended model introduced in this work,} we aim to characterize the influence of the FLR on the appearance of STBs and the chaotic transport of guiding-center test particles. We consider a non-monotonic radial equilibrium electric field profile and an electrostatic perturbation regarded as the superposition of coherent harmonic waves traveling in the poloidal and toroidal directions. Additionally, monotonic profiles for the safety factor and the toroidal velocity are assumed. Since the applied model has a Hamiltonian structure, the phase space flow generated by solving the equations of motion is area-preserving in an adequate Poincaré surface of section.\par 

The Larmor radius and the electrostatic perturbation amplitude of one harmonic mode are taken as control parameters to be varied. We introduced suitable quantifiers for the chaotic transport like the transmissivity of the barrier and the transport current, associated with the probability of a given orbit to escape and the escape rate \textcolor{black}{of orbits}, respectively. Bifurcation diagrams of the STB as a function of the control parameters are obtained\textcolor{black}{, indicating the relationship between the FLR and the phase space topology.}\par 

This paper is organized as follows: Section \ref{sec:model} describes the drift wave \textcolor{black}{guiding-center test particle} transport model with finite Larmor radius effect. Section \ref{sec:STB} presents a dynamical analysis of the phase space of our numerical map, emphasizing the computation of the STB and presenting bifurcation diagrams for the STB as a function of the perturbation amplitude for different Larmor radii. The diagnostics used to characterize the reduction of chaotic transport are introduced in Section \ref{sec:transport}, as well as a comprehensive analysis of the corresponding parameter plane. Our conclusions are left to Section \ref{sec:conclusions}. 

\section{Drift wave test particle transport model}\label{sec:model}

\begin{figure}[b]
    \centering
    \includegraphics[width = 0.47\textwidth]{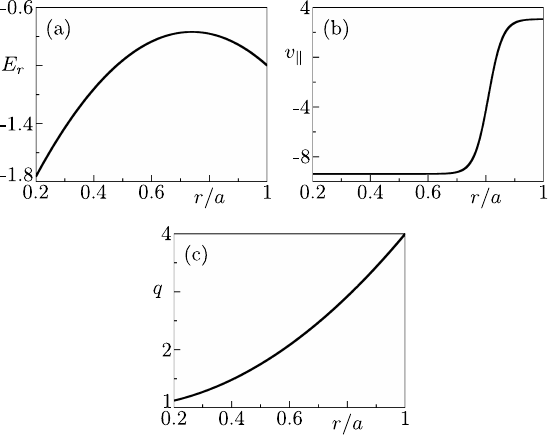}
    \caption{Dimensionless plasma profiles, mainly, (a) the radial electric field, $E_r(r)$, (b) the parallel velocity, $v_\parallel(r)$, and (c)~the safety factor, $q(r)$.}
    \label{fig:plasma_profiles}
\end{figure}

\begin{figure}[t]
    \centering
    \includegraphics[width = 0.47\textwidth]{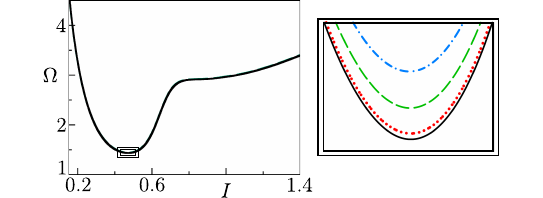}
    \caption{Rotation number profile for the integrable case, \mbox{$H_1 = 0$}. The black (solid) curve corresponds to the massless approximation, $\rho = 0.0$, the red (dotted) one to \mbox{$\rho = 1.637\times 10^{-2}$}, the green (dashed) one to  $\rho = 3.770\times 10^{-2}$, and the blue (dashed-dotted) one to $\rho = 5.556\times 10^{-2}$.}
    \label{fig:integrable_rotation_number}
\end{figure}

\begin{figure*}
    \centering
    \includegraphics[width = \textwidth]{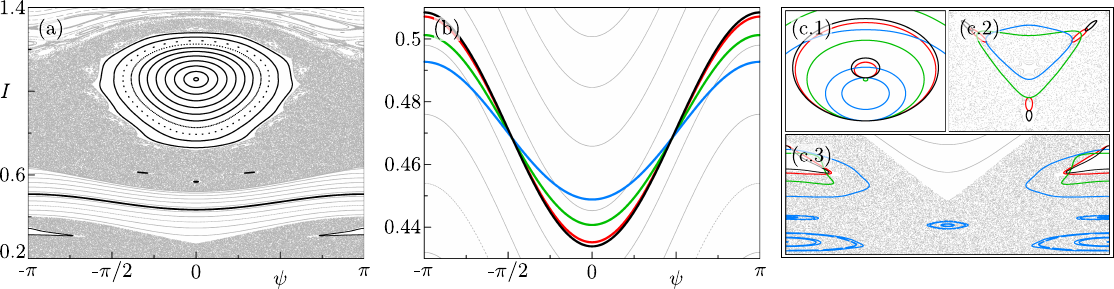}
    \caption{Phase space for $\phi_1 = 0.0$. In gray and black colors, we show chaotic and quasi-periodic orbits for the massless case, $\rho = 0.0$. Red, green, and blue colors correspond to orbits for $\rho = 1.637\times 10^{-2}$, $\rho = 3.770\times 10^{-2}$ and $\rho = 5.556\times 10^{-2}$, respectively. Panel (a) shows only the Poincaré section for $\rho = 0.0$; meanwhile, panels (b) and (c) are magnifications of the shearless curve and the main islands, respectively, including the FLR effect. Specifically, in panel (c), we show magnifications around the black-colored islands in panel (a), ordering them from the top to the bottom.}
    \label{fig:phase_space_phi1=0}
\end{figure*}

    \textcolor{black}{The model presented in this section builds on the drift wave test particle transport framework introduced in Ref.~\onlinecite{horton1998}, with an extension that accounts for the Larmor radius effect that arises from a non-uniform electric field. This modification enables us to study some interesting chaotic transport features in a more realistic situation in which the transport behavior of test particles varies based on their specific Larmor radii.}\par 
    
    It is considered a test particle that is immersed in a large aspect-ratio tokamak plasma, i.e.,\ \mbox{$1/\varepsilon = R/a \gg 1$}, with $R$ and $a$ the major and minor radii of the plasma column, respectively. The particle's guiding center is moving along the lines of the magnetic field, $\mathbf{B}(\mathbf{x})$, with velocity $\mathbf{v}_\parallel(\mathbf{x})$ and drifted by the gyro-averaged velocity $\mathbf{v}_\mathbf{E}(t,\mathbf{x})$, evaluated at the \mbox{guiding center}, so that  

\begin{subequations}\label{eq:original_motion}
    \begin{equation}\label{eq:guiding_center_motion}
\dfrac{\mathrm{d} \mathbf{x}}{\mathrm{d} t} = \mathbf{v}_\parallel + \mathbf{v}_\mathbf{E},
\end{equation}
\begin{equation}\label{eq:drift_velocity}
\mathbf{v}_\mathbf{E} = \left(\mathbf{E}+\dfrac{1}{4}\rho^2\nabla_\perp^2\mathbf{E}\right)\times\dfrac{\mathbf{B}}{B^2},
\end{equation}
\end{subequations}

    \noindent where $\rho$ is the Larmor radius, $\mathbf{E}(t,\mathbf{x})$ is the electric field and $\nabla_\perp^2$ is the Laplacian taken on the perpendicular plane to $\mathbf{B}$. \textcolor{black}{It is assumed that test particles are sufficiently diluted so as not to affect the electric and magnetic fields.}\par 
    
    The gyro-averaged drift velocity results from the averaging method presented in Ref.~\onlinecite{hazeltine2018}, taking up to second-order space-varying electric field contributions. \textcolor{black}{Where $\partial E / \partial t \ll \omega_\mathrm{g} E$, with $\omega_\mathrm{g}$ the cyclotron angular frequency, or gyro-frequency, of the particle}.  This is a usual approximation that only considers the influence of a non-uniform electric field, and plays a fundamental role in describing transport in the presence of fast particles, such as alpha particles in burning plasmas \cite{manfredi1997, annibaldi2002}.\par 

    Although other drift velocities\textcolor{black}{, such as the $\mathbf{B}\times\nabla B$ drift,} could be considered to evaluate the guiding center orbit due to the finite Larmor radius (FLR) effect, \textcolor{black}{we limit our analysis for simplicity. As we will show, this choice is justified by the fact the drift velocity in \eqref{eq:drift_velocity} does not affect the Hamiltonian nature of the original dynamical system.}\par 
    
    Hence, it is considered a magnetic field such that

\begin{equation}
  \mathbf{B}(r) = B_\theta(r)\hat{e}_\theta + B_\varphi(r)\hat{e}_\varphi,  
\end{equation}

    \noindent with $B\approx B_\varphi \gg B_\theta$ and $B \approx B_0$, where $B_0$ is constant; and where $r$, $\theta$ and $\varphi$ are the radial, poloidal and toroidal coordinates of the toroidal system, respectively.\par
    
    Furthermore, the electric field is considered as a rotation-free vector field, $\nabla\times\mathbf{E} = \mathbf{0}$.  In equilibrium, it is completely described by the radial component $E_r(r)\hat{e}_r$. In non-equilibrium, \textcolor{black}{a simplified model of drift wave transport is adopted, incorporating electrostatic potential fluctuations, $\tilde{\phi}(t,\mathbf{x})$, characterized by a single dominant spatial mode and harmonics of the lowest dominant angular frequency, $\omega_0$, in the drift wave spectrum \cite{brower1985, horton1998, marcus2019}. The interaction between the components of the electric field is not considered in this model}. Therefore,

\begin{subequations}\label{eq:electric_field}
\begin{equation}
\mathbf{E}(t,\mathbf{x}) = E_r(r)\hat{e}_r - \nabla \tilde{\phi}(t,\mathbf{x}),
\end{equation}
\begin{equation}\label{eq:perturb_potential}
\tilde{\phi}(t,\theta, \varphi) = \sum_{n} \phi_{n} \cos(M\theta-L\varphi-n\omega_0 t + \alpha),
\end{equation}
\end{subequations}

    \noindent where $M$ and $L$ are dominant wave numbers in the poloidal and toroidal directions, respectively, $\phi_n$ the wave amplitude of each mode and $\alpha$ a constant phase.\par 
    
    Since $\mathbf{B}$ is essentially toroidal, we only look at the Laplacian on the $(r,\theta)$ plane. Then, by introducing two new variables, namely the action, $I$, and the angle, $\psi$, defined as

\begin{subequations}\label{eq:action-angle}
\begin{equation}
I = \left(\dfrac{r}{a}\right)^2,
\end{equation}
\begin{equation}
\psi = M\theta-L\varphi,
\end{equation}
\end{subequations}

    \noindent and performing an adimensionalization by using the characteristic scales $a$, $E_a = \vert E_r(I=1) \vert$ and $B_0$, according to relations

\begin{equation} \label{eq:adimensionalization}
\begin{split}
   E_r'&=\dfrac{E_r}{E_a}, \quad \phi_n'=\dfrac{\phi_n}{aE_a}, \quad v_\parallel'=\dfrac{B_0}{E_a}v_\parallel,\quad t'=\dfrac{E_a}{aB_0} t,\\
   \omega_0' & =\dfrac{aB_0}{E_a}\omega_0, \quad \rho'=\dfrac{\rho}{a}, 
\end{split}
\end{equation} 

    \noindent the equations of motion \eqref{eq:guiding_center_motion} simplify into the 1.5-degrees-of-freedom dynamical system  

\begin{subequations} \label{eq:model}
\begin{equation}\label{eq:action}
\dfrac{\mathrm{d} I}{\mathrm{d} t'} = - f(I)\dfrac{\partial \tilde{\phi}'(t',\psi)}{\partial \psi}, 
\end{equation}
\begin{equation}\label{eq:angle}
\dfrac{\mathrm{d} \psi}{\mathrm{d} t'} = g(I)  +  \dfrac{\mathrm{d} f(I)}{\mathrm{d} I} \tilde{\phi}'(t',\psi),
\end{equation}
\end{subequations}

    \noindent where $f$ and $g$ are functions of $I$, and $g$ is called the twist function of the system. This function is important in the KAM theorem, as the twist condition of the system (non-degeneracy condition), ${\mathrm d} g / {\mathrm d} I \neq 0$ for all $I$, ensures the robustness of KAM tori under perturbations\cite{lichtenberg1992}.  Thus, a challenging dynamics, related to chaotic transport, occurs near the invariant curve  where ${\mathrm d} g / {\mathrm d} I = 0$, i.e.,\ where the twist condition is broken\cite{howard1984}. In our model, the functions $f$ and $g$, dropping the prime notation henceforth, correspond to\par

\begin{subequations}\label{eq:aux_functions}
\begin{equation}
f(I) = 2M\left(1-\dfrac{\rho^2M^2}{4I}\right), 
\end{equation}
\begin{equation}\label{eq:winding_function}
g(I)= \varepsilon v_\parallel(I)\dfrac{M-Lq(I)}{q(I)}-\dfrac{M}{\sqrt{I}}\left(1+\rho^2\Delta\right) E_r(I),
\end{equation}
\end{subequations}

    \noindent where $\Delta = {\mathrm{d}}/{\mathrm{d}I}\left(I\,{\mathrm{d}}/{\mathrm{d}I}\right)-{1}/{(4I)}$ is an operator which we introduce here and $q(I) = \varepsilon\sqrt{I}B_\varphi/B_\theta$ is the safety factor profile. 

    Notice that, since $\partial \dot{I}/\partial I + \partial \dot{\psi}/\partial \psi = 0$, where the dot notation is the total time derivative, there is a Hamiltonian function, $H(t,\psi, I)$, such that 

\begin{equation}
    \dot{\psi} = \dfrac{\partial H}{\partial I}, \quad \dot{I} = -\dfrac{\partial H}{\partial \psi}, 
\end{equation}

    \noindent which can be decomposed into an integrable, $H_0(I)$, and a perturbative part, $H_1(t,\psi,I)$,  

\begin{equation}
    H(t,\psi,I) = H_0(I) + H_1(t,\psi,I),
\end{equation}

    \noindent where

\begin{subequations}\label{eq:Hamiltonians}
\begin{equation}
H_0(I) = \int^I g(I')\, \mathrm{d}I', 
\end{equation}
\begin{equation}
H_1(t,\psi,I)= f(I)\tilde{\phi}(t,\psi).
\end{equation}
\end{subequations}

    Thus, when $H_1 = 0$, i.e., \ when $\phi_n = 0$ for all modes, $I$ remains constant, and the guiding center orbit traces a helix of constant radius along a curve on the same equilibrium magnetic surface. On the other hand, when $H_1 \neq 0$, the integrability of the system is broken, leading to chaotic behavior and chaotic transport.\par

    In the limit when $\rho = 0$, the dynamical system \eqref{eq:model} reduces to the same equations \textcolor{black}{of the original model introduced} in Ref.~\onlinecite{horton1998}. For these, it has been shown that non-monotonic profiles of $E_r(I)$, $q(I)$ and $v_\parallel(I)$ can lead to a non-twist behavior, \mbox{$\mathrm{d}g/\mathrm{d}I = 0$}, and the emergence of \textcolor{black}{a special type of transport barrier that reduces the chaotic transport of particles, usually called shearless transport barrier (STB)\cite{rosalem2014,marcus2019,grime2022shearless,osorio2023ExB}}.\par 
    
    However, we must be careful because the massless approximation is violated for cases when particles are fast, e.g., \ alpha particles, or even if we study impurity transport since the impurity temperature can be equivalent to the plasma's\cite{kim1994}. In those cases, $\rho\approx0$ is not valid anymore.  \par
    
    We aim to investigate the influence that the FLR effect has on the onset of such barriers and also on chaotic transport. We consider for this a non-monotonic radial equilibrium electric field profile, and monotonic profiles for the safety factor and the parallel velocity.\par  
      
\section{influence of the FLR on shearless transport barriers}\label{sec:STB}

\begin{figure}[b]
    \centering
    \includegraphics[width = 0.42\textwidth]{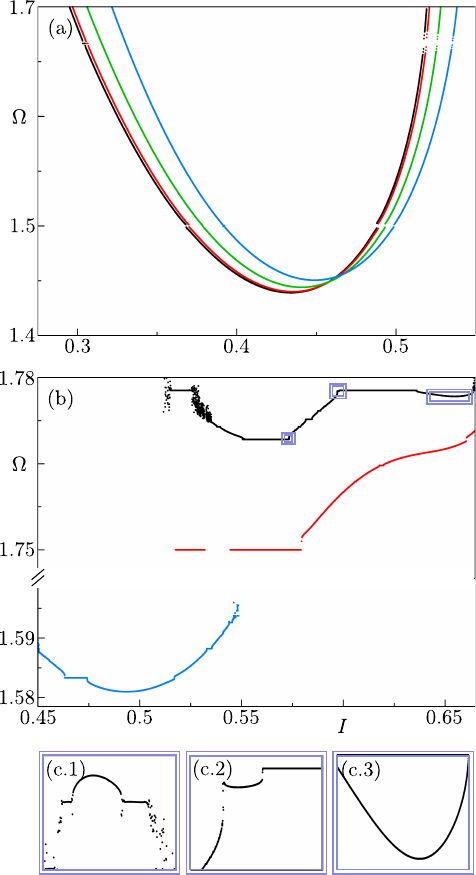}
    \caption{Rotation number profiles for $H_1\neq0$, (a) $\phi_1=0.0$, and (b) $\phi_1=2.283\times 10^{-2}$. Black, red, green, and blue colors correspond to $\rho = 0.0$, $\rho = 1.637\times 10^{-2}$, $\rho = 3.770\times 10^{-2}$ and $\rho = 5.556\times 10^{-2}$, respectively. For the black-colored profile in panel (b), we present in panel (c) magnifications around three different extrema where shearless curves can be identified. Specifically, (c.1) is near $I = 0.57$, (c.2) near \mbox{$I = 0.60$} and (c.3) near $I = 0.65$.}
    \label{fig:rotation_number}
\end{figure}

\begin{figure*}
    \centering
    \includegraphics[width = 0.66\textwidth]{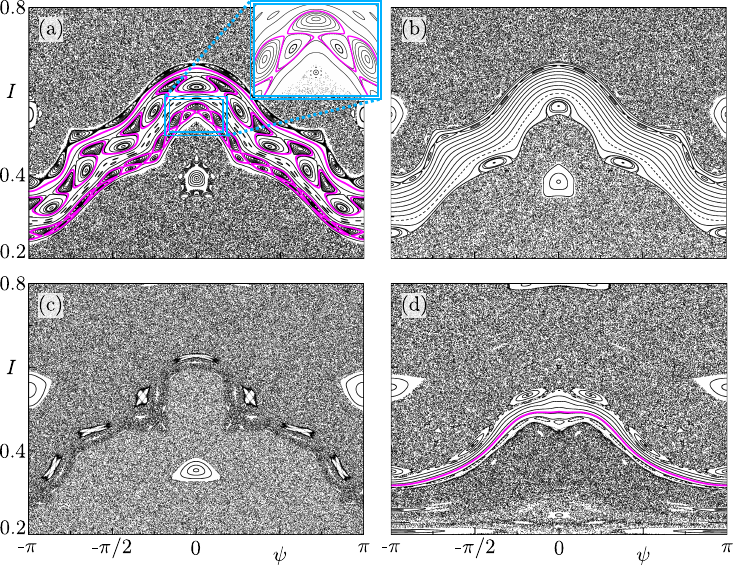}
    \caption{Poincaré sections for $\phi_1=2.283\times 10^{-2}$ and (a) $\rho = 0.0$, (b) $\rho = 1.637\times 10^{-2}$, (c) $\rho = 3.770\times 10^{-2}$ and \mbox{(d) $\rho = 5.556\times 10^{-2}$}. Shearless transport barriers are colored in magenta.}
    \label{fig:phase_space_phi1=0.009345}
\end{figure*}

    Non-twist behavior is found in many physical problems, particularly in plasmas and fluid dynamics \cite{oda1995, morrison2000,delCastillo1993Rossby,balescu1998,behringer1991}. Systems exhibiting that behavior present resilient barriers that inhibit chaotic transport \cite{delCastillo1996}, also known as shearless transport barriers (STBs), and feature a characteristic dynamics since the KAM theorem is not applicable due to the degeneracy of the system itself \cite{howard1995}.\par 
    
    Such barriers and the neighboring KAM curves, composing the non-twist barrier, are expected to be the latest invariant curves to be destroyed and also the easiest to restore \cite{shinohara1997}. Moreover, after the breakup, the remnant of the non-twist barrier, often accompanied by sticky behavior, acts as a partial barrier to transport\cite{szezech2009}. So, the control of the chaotic transport of particles is generally reduced to knowing whether the STB exists, how robust it is, and even more, how resilient and effective the partial barriers are. Specifically, in this section, we are interested in evaluating the influence of the FLR on the onset of STBs as the control parameters vary.\par     
    For that, let us establish the numerical map \mbox{$\mathbf{z}_{j+1} =  F(\mathbf{z}_j)$}, where \mbox{$\mathbf{z}_j=(\psi_j, I_j) = (\psi(t_j), I(t_j))$}, \mbox{$j \in \mathbb{N}$} and $F(\mathbf{z}_j)$ is a numerical integration of the dynamical system \eqref{eq:model} that evolves the orbit, given the initial condition $\mathbf{z}_0$ at time $t_0=0$, from $t_j$ to $t_j+T$, with $T = 2\pi/\omega_0$. By doing that, we construct stroboscopic Poincaré sections of the periodic, quasi-periodic, or chaotic guiding-center orbits. The results presented in this work were obtained using the numerical integrator Runge-Kutta-Dormand-Prince of 8(7) order \cite{engeln1996},  \textcolor{black}{with an error tolerance of $10^{-13}$. In particular, adaptive step-size Runge-Kutta methods provide efficient performance while maintaining acceptable error levels, even when compared to symplectic integrators, depending on the specific problem\cite{lazarotto2024}.} 

    So, to find a STB, we calculate the rotation number profile $\Omega(\mathbf{z})$ and identify whether it has an extreme value \mbox{$(\partial\Omega/\partial I)_{\mathbf{z}_{\mathrm{STB}}} = 0$}, from which a shearless orbit, corresponding to the barrier, can be generated by using $\mathbf{z}_{\mathrm{STB}}$ as the initial condition.  The rotation number of an orbit is essentially the average angular displacement experienced by the orbit, so it will be rational if the orbit is periodic and irrational if it is quasi-periodic. To calculate it, we use the method proposed in Ref.~\onlinecite{das2018}, by which we obtained a reliable convergence of $\Omega$ with less iterations. Therefore,

\begin{subequations} \label{eq:rotation_number}
    \begin{equation}\label{eq:rotation_number_average}
        \Omega = \dfrac{1}{2\pi}\sum_{j=0}^{K-1}\hat{s}_{j,K}\Pi(F(\mathbf{z}_{j})-\mathbf{z}_j), 
    \end{equation}
    \begin{equation}\label{eq:rotation_number_weighter}
        \hat{s}_{j,K} = \dfrac{s(j/K)}{\sum_{j=0}^{K-1}s(j/K)},
    \end{equation}
    \begin{equation}\label{eq:superconvergent_weighted_average}
        s(x) = \left\{ \begin{array}{lcc}
			\exp\left(\dfrac{-1}{x(1-x)}\right), & \mathrm{for} & x \in (0,1) \cr
                && \cr
			0, & \mathrm{for} & x \notin (0,1),
		\end{array}
		\right.
    \end{equation}
\end{subequations}

    \noindent where $\Pi$ is a suitable angular projection, that for our particular problem can be taken as $\Pi (\mathbf{z}_j) = \psi_j$. Also, notice we are performing a normalization by $2\pi$. \textcolor{black}{In \mbox{appendix \ref{appendix:convergence}}, we compare the convergence of the rotation number using an equal-weighted average, $s(x) = 1$, which is a common approximation,  with the super-convergent method that weights the average according to the relation \eqref{eq:superconvergent_weighted_average}.}\par 

    For the integrable case, $H_1 = 0$, it is easy to show the rotation number of the $T$-period stroboscopic map does not depend on the initial angle $\psi_0$ and is equal to \mbox{$\Omega = g(I)/\omega_0$}. In consequence, according to the configuration of the profiles $E_r(I)$, $v_\parallel(I)$ and $q(I)$, see equation \eqref{eq:winding_function}, the twist condition can be violated and, therefore, STBs appear at the zero-derivative points of $g(I)$.\par  

    Specifically, we take into account the plasma profiles and parameters for the tokamak TCABR \cite{marcus2008,severo2009,galvao2015}, for which the minor and major radii are \mbox{$a=0.18$ m} and \mbox{$R = 0.61$ m}, respectively, and the toroidal magnetic field is \mbox{$B_0 = 1.20$ T}. The plasma profiles are taken in the form

    \begin{subequations} \label{eq:plasma_profiles}
    \begin{equation}
        E_r(r) = E_0 + E_1\left(\dfrac{r}{a}\right) + E_2 \left(\dfrac{r}{a}\right)^2, 
    \end{equation}
    \begin{equation}
        v_\parallel(r) = v_0 + v_1 \tanh \left[\beta_1 \left(\dfrac{r}{a}\right)+\beta_0\right],
    \end{equation}
    \begin{equation}
        q(r) = \left\{ \begin{array}{lcc}
			q_0 + (q_a-q_0)\left(\dfrac{r}{a}\right)^2, & \mathrm{for} & r \leq a \cr
                && \cr
			q_a\left(\dfrac{r}{a}\right)^2, & \mathrm{for} & r > a,
		\end{array}
		\right.
    \end{equation}
\end{subequations}
    
    \noindent as shown Figure~\ref{fig:plasma_profiles}. For these profiles, \textcolor{black}{which have already been examined in previous works\cite{marcus2019,grime2022shearless},} we consider the dimensionless parameters:  \mbox{$\beta_0 = -16.42$}, \mbox{$\beta_1 = 20.30$}, \mbox{$q_0 = 1.0$} and \mbox{$q_a = 4.0$}; and, before carrying out the adimensionalization, the parameters: \mbox{$v_0 = -5.98$ km/s}, \mbox{$v_1 = 11.793$ km/s}, \mbox{$E_0 = -6.0$ kV/m}, \mbox{$E_1 = 5.751$ kV/m} and \mbox{$E_2 = -2.592$ kV/m}, i.e.,\ \mbox{$E_a = 2.274$ kV/m}. Furthermore, for the electrostatic potential perturbation, we employ as dominant spatial modes, \mbox{$M = 16$} and \mbox{$L = 3$}, and as fundamental angular frequency \mbox{$\omega_0 = 60.0$ rad/ms} (approximately \mbox{$5.70$ rad} after the adimensionalization).\par
    
    As a result, for the integrable case, non-monotonic behavior in the rotation number profile is obtained, see \mbox{Figure~\ref{fig:integrable_rotation_number}}, and, consequently, a STB is expected to exist. We show in the right panel of the figure a magnification localized in the reversed-shear region, where a subtle difference in the profile can be observed by varying the Larmor radius. For this case, the shearless point position displacement and its rotation number vary slightly with $\rho$. Because of that, near the STB, some rational and irrational orbits are not accessible anymore \textcolor{black}{for some particles with large Larmor radius}, if compared with the former case $\rho = 0$.\par 

    Now, concerning $H_1 \neq 0$ for the non-integrable scenario, four harmonics corresponding to the main resonances, $n=2,3,4$, and the non-resonant mode, \mbox{$n=1$}, according to Figure~\ref{fig:integrable_rotation_number}, are taken. For the resonant modes, we adopt the amplitudes \mbox{$0.80$ V}, \mbox{$1.50$ V} and \mbox{$0.85$ V}, which become the dimensionless fixed parameters \mbox{$\phi_2=1.95\times 10^{-3}$}, \mbox{$\phi_3=3.66\times 10^{-3}$} and \mbox{$\phi_4=2.08\times 10^{-3}$}, respectively. Additionally, we regard the amplitude $\phi_1$, which corresponds to the non-resonant mode, as a control parameter. Some studies have shown that the STB can be repeatedly destroyed and restored as the amplitude of non-resonant modes varies \cite{marcus2019}, \textcolor{black}{while resonant modes are associated with high transport coefficients \cite{marcus2008}. Accordingly, we establish the fluctuation level linked to the resonant modes, which induces a degree of chaotic transport, and examine the influence of non-resonant modes on regulating this transport in conjunction with the finite Larmor radius.}  

    Thus, for the scenario $\phi_1 = 0$ shown in Figure~\ref{fig:phase_space_phi1=0}, the impact of the FLR effect is illustrated. The left panel provides an overview of the Poincaré section for the massless case, where most of the KAM tori are broken. Only the non-twist barrier, formed by the STB (colored in black) and the neighboring invariant curves, survives. The effect of introducing the FLR is shown in the panels (b) and (c). Panel (b) presents a magnification of the STB region, while panel (c) shows some islands. The orbits in gray and black, which correspond to $\rho=0$, are included as background and reference. \par 

    The barrier in Figure~\ref{fig:phase_space_phi1=0}(b) exhibits minimal variation with $\rho$, except for a radial displacement and a subtle difference in shape. This can also be verified by looking at the rotation number profile in Figure~\ref{fig:rotation_number}(a). For this perturbation scenario, most differences are evidenced by looking at the islands. For example, as $\rho$ increases, in panel~\ref{fig:phase_space_phi1=0}(c.1), the center of the main island, close to \mbox{$I = 1.0$}, shifts; in panel~\ref{fig:phase_space_phi1=0}(c.2), a local bifurcation occurs for the periodic orbits near $I = 0.6$; and, in panel~\ref{fig:phase_space_phi1=0}(c.3), new island chains appear in the lower chaotic region for the largest value considered of $\rho$. Also, a variation in the size of the islands can be observed. \par 

    When we increase the value of $\phi_1$, in contrast to the previous results, various transport barrier scenarios are obtained by varying $\rho$. This is depicted in the Poincaré sections of Figure~\ref{fig:phase_space_phi1=0.009345} for  $\phi_1 = 2.283\times 10^{-2}$. In these scenarios, we observe that a bifurcation of the shearless curve occurs for the massless approximation, leading to the appearance of three different shearless transport barriers, see \mbox{panel (a)}. This bifurcation can emerge due to cubic and quartic contributions in non-twist maps  \cite{howard1995,grime2023}. Then, introducing the FLR effect, panels (b) and (c), the STBs disappear, leaving invariant curves and a partial barrier with sticky behavior, respectively. Eventually, a single shearless curve arises again when increasing the FLR until $\rho = 5.556\times 10^{-2}$, as shown in (d).   The rotation number profiles of the previous cases are shown in Figure~\ref{fig:rotation_number}(b) and \ref{fig:rotation_number}(c); only for \mbox{$\rho= 3.770\times 10^{-2}$}, there is no profile because all KAM tori are broken. \textcolor{black}{These results indicate that, for the same fluctuation levels, while some test particles experience reduced transport due to the presence of transport barriers, others do not exhibit the same resistance to transport. This is beneficial, as it may provide a selective decontamination mechanism for specific particles.} \par

   To get an overall view of how $\phi_1$ affects the existence of shearless curves under different scenarios of $\rho$, we construct STB bifurcation diagrams as shown in \mbox{Figure~\ref{fig:bifurcation_diagram}}. For clarity, panel (a) includes only the massless case (black) and the $\rho = 1.637 \times 10^{-2}$ case (red), while \mbox{panel (b)} shows scenarios for $\rho = 3.770 \times 10^{-2}$ (green) and $\rho = 5.556 \times 10^{-2}$ (blue). Fundamentally, for each value of $\phi_1$, we compute the rotation number profile and examine it to identify the presence of extreme points. If they are found, we plot points corresponding to the rotation numbers, $\Omega_{\mathrm{STB}}$, of the shearless orbits.\par 

    Particularly for the parameters and profiles we have chosen, \textcolor{black}{bifurcations of the shearless curve} are inhibited as the FLR effect increases. They are common for large non-resonant perturbation amplitudes, for which cubic and quartic contributions appear to gain relevance in the numerical map\cite{howard1995, grime2023}. Furthermore, it is interesting to remark that the STB becomes more resilient to small and medium perturbations as $\rho$ increases. This is clearly evidenced in panel (b), where the first two intervals of the barrier are larger and have almost no gaps compared to panel (a). In general, for the largest value of $\rho$ considered, represented by the blue-colored bifurcation diagram, the barrier is broken up less frequently and restored more easily.\par 
    
    \textcolor{black}{These results highlight the role of the FLR in influencing the phase space topology and controlling chaotic transport. The shearless barrier bifurcation diagrams identify specific parameter intervals where significant transport and nonlinear mechanisms occur, including STB break-up, multiple-separatrix reconnection, and STB reemergence. These mechanisms are significant not only from a dynamical systems perspective but also for plasma physics, as highlighted in other studies \cite{escande2016}.}  
   
\section{Chaotic transport reduction}\label{sec:transport}

\begin{figure}[t]
    \centering
    \includegraphics[width = 0.44\textwidth]{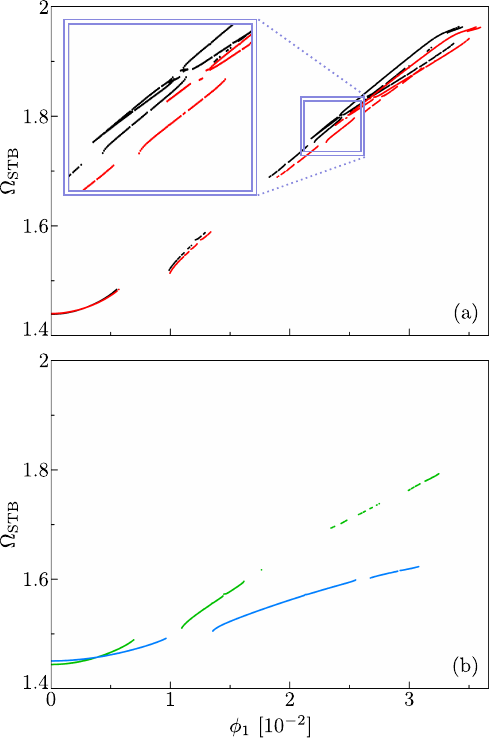}
    \caption{Bifurcation diagrams of the shearless transport barrier for (a) $\rho=0$ (black) and $\rho=1.637\times 10^{-2}$ (red), and for (b) $\rho = 3.770\times 10^{-2}$ (green) and $\rho = 5.556\times 10^{-2}$ (blue).}
    \label{fig:bifurcation_diagram}
    \end{figure}

\begin{figure}[t]
    \centering
    \includegraphics[width = 0.46\textwidth]{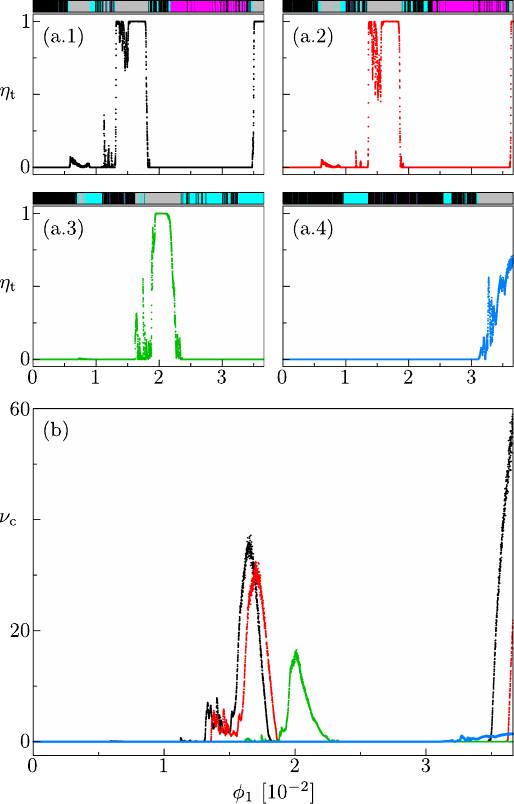}
    \caption{Diagrams with $\phi_1$ of the (a) transmissivity and (b) transport current for \mbox{$\rho=0$} (black), \mbox{$\rho=1.637\times 10^{-2}$} (red), \mbox{$\rho = 3.770\times 10^{-2}$} (green) and \mbox{$\rho = 5.556\times 10^{-2}$} (blue). The color bars in the panels (a) represent the $\phi_1$ intervals in which it was identified one shearless curve (black), two or more shearless \mbox{curves (magenta)}, zero transmissivity but no STB (cyan), and $\eta_\mathrm{t} > 0$ (gray).}
    \label{fig:diagrams}
\end{figure}

   In this section, we explore additional diagnostics to gain broader insights into the influence of the FLR effect.  We focus on chaotic transport's behavior as the control parameters vary, particularly after the non-twist barrier is broken.\par 
   So, we analyze the transmissivity, $\eta_\mathrm{t}$, and the ``transport current'', $\nu_\mathrm{c}$, which are measurements compared to the probability of a given chaotic orbit escape from some region to another one and the escape rate of orbits, respectively. They are computed as follows:

   \begin{itemize}
       \item We select an ensemble of $N$ randomly chosen initial conditions, $\{\mathbf{z}_0^i\}_{i=1}^N$, in a small chaotic area under the non-twist barrier. A previous survey of the phase space must be done to guarantee that by varying the parameters, the orbits are still in a chaotic region. 
       \item We integrate each orbit until either a maximum of $K$ iterations or it crosses the threshold $I_\eta$, i.e.,\ $I_j^i > I_\eta$. If the second criterion is fulfilled, we record the time $K_\mathrm{c}^i = j$, and the orbit identified by $\mathbf{z}_0^i$ is counted as an escaping orbit. Otherwise, $K_\mathrm{c}^i = 0$. By doing this, we calculate a mean escape time only considering the orbits that actually escape, as shown next.      
       \item We compute the transmissivity as \mbox{$\eta_\mathrm{t} = N_\mathrm{t}/N$}, where $N_\mathrm{t}$ is the total number of escaping orbits, and the transport current as $\nu_\mathrm{c} = \eta_\mathrm{t} / (\langle K_\mathrm{c} \rangle / K)$, where $\langle K_\mathrm{c} \rangle = \sum_{i=1}^N K_\mathrm{c}^i / N_\mathrm{t}$ is the mean escape time.  
   \end{itemize}

   Notice that if $\eta_\mathrm{t} > 0$, at least one orbit of the ensemble escapes; therefore, the non-twist barrier does not exist. Conversely, if $\eta_\mathrm{t} = 0$, it is very likely that at least one KAM torus survives and acts as a barrier to transport. A small transmissivity indicates resistance to transport, such as sticky behavior or remnants of the non-twist barrier, as shown in Figure \ref{fig:phase_space_phi1=0.009345}(c). This resistance prevents a fraction of the particles from crossing within the time period $K$. On the other hand, large values of $\eta_\mathrm{t}$ imply that the characteristic escape time of the particles is less than $K$.\par 

   However, equal $\eta_\mathrm{t}$ scenarios do not translate into equal chaotic transport conditions. This is because $\langle K_\mathrm{c} \rangle$ can vary significantly between scenarios. High transport situations occur when the transmissivity is large, and the mean escape time is small, i.e., when the transport current $\nu_\mathrm{c}$ is large. Conversely, low transport occurs when the transport current is small, which corresponds to scenarios where $\langle K_\mathrm{c} \rangle$ is large, or $\eta_\mathrm{t}$ is small, or both.\par 

\begin{figure}[htb]
    \centering
    \includegraphics[width = 0.47\textwidth]{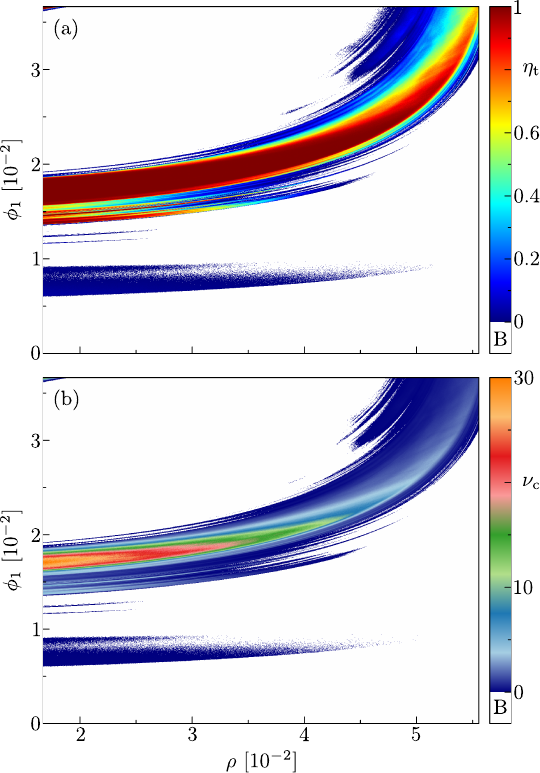}
    \caption{Parameter spaces with $\rho$ and $\phi_1$. Panel (a) shows the transmissivity, $\eta_\mathrm{t}$, and panel (b) the current, $\nu_\mathrm{c}$. The white color indicates, in both panels, that some type of barrier exists \mbox{(B labeled scenario in the color bar)}.}
    \label{fig:parameter_space}
\end{figure}

   In Figure \ref{fig:diagrams}, we present diagrams of the transmissivity and the transport current as a function of $\phi_1$. These diagrams were obtained using $N = 10^3$ randomly chosen initial conditions in the intervals $\psi_0 = [-0.54, -0.42]$ and $I_0 = [0.275, 0.29]$, integrated until either $K = 5 \times 10^3$ crossings in the Poincaré section or until $I_j^i > 0.8$. Moreover, from the results in Figure \ref{fig:bifurcation_diagram}, we identify intervals of $\phi_1$ where scenarios with or without shearless transport barriers occur. Combining these results, we provide a detailed picture of the control of chaotic transport for the four Larmor radius values considered.\par

   Then, from panel \ref{fig:diagrams}(a), we can say, for practical purposes, that there are similar behavior patterns in the first three cases, except for a shift in the value of the perturbation at which maximum transmissivity occurs. In contrast, for the last case, the transmissivity never reaches the maximum value, $\eta_\mathrm{t} = 1$, and increases more slowly with $\phi_1$. Of the four cases, this one is the most conclusive in terms of chaotic transport reduction through transmissivity diagnosis. \par 
   
   Additionally, as pointed out in the previous section, the color bars show that the STB typically bifurcates for small $\rho$ values and large perturbations,  and the first intervals of $\phi_1$ are larger as $\rho$ increases. Nonetheless, we now observe new intervals in which no STB exists, but different types of barriers inhibit chaotic transport, such as those in Figures \ref{fig:phase_space_phi1=0.009345}(b) and \ref{fig:phase_space_phi1=0.009345}(c). \textcolor{black}{Additional mechanisms for transport reduction are likely to appear, such as those discussed in Refs.~\onlinecite{falessi2015} and~\onlinecite{di2018}.}  As shown in panels (a.3) and (a.4), these barriers appear more frequently for the largest values of $\rho$. \par 

   In panel \ref{fig:diagrams}(b), it becomes clearer that the FLR effect leads to a reduction in chaotic transport. As $\rho$ increases, the transport current consistently decreases, with its peak shifting towards larger values of $\phi_1$. Furthermore, while the black, red, and green cases have a maximum transport current on the order of $10^1$, the blue case reduces the transport current by one order of magnitude.\par 
   
   In a combined manner, the diagnoses of transmissivity and current transport allow us to discern subtle differences between scenarios. This is generally illustrated in the $\rho\times\phi_1$ parameter spaces shown in Figure \ref{fig:parameter_space}. There, scenarios with no chaotic transport are depicted in white, where STBs, KAM tori, or strong partial barriers may emerge. Conditions with low chaotic transport are predominantly shown in blue, indicative of partial barriers as well. Then, we observe that transport barriers tend to be more robust as the Larmor radius increases, requiring larger values of $\phi_1$ to maximum transport. Moreover, following the main stripe, while we cannot observe significant variation in transmissivity, there is a clear systematic decrease in transport current conditions as $\rho$ increases. \par

  Finally, the lower stripe, which appears for most of the Larmor radius interval and within a perturbation interval for $\phi_1$ of \mbox{$0.5\times10^{-2}$} to $1\times10^{-2}$, is linked to an interesting partial barrier dynamics, as detailed in Ref.~\onlinecite{mugnaine2024}. In these scenarios, even-period twin islands exhibit stable/unstable manifolds of the associated hyperbolic points that share a common branch which, in turn, separates two chaotic regions. Consequently, very few orbits manage to cross.\par   
  
   In conclusion, our findings suggest that the FLR effect plays a crucial role in promoting more robust transport barriers and reducing the chaotic transport \textcolor{black}{of test particles}. We encourage other researchers to explore further aspects of this work, particularly the influence of the radial equilibrium electric field profile in chaotic transport and the symplectic map developed in \textcolor{black}{Appendix \ref{appendix:map}}.    
   
\section{Conclusions}\label{sec:conclusions}

A drift-wave guiding-center test particle transport model has been implemented to evaluate the influence of the Larmor radius on chaotic transport and the onset of transport barriers in tokamaks. Specifically, we examined the guiding-center motion of a test particle as it moves along the magnetic field lines and is drifted by a gyro-averaged velocity caused by a non-uniform electrostatic field. The numerical simulations presented in this work were performed using parameters and plasma radial profiles for the tokamak TCABR; nonetheless, the results are valid for a wide class of \textcolor{black}{magnetic confinement devices, such as large aspect-ratio tokamaks and Helimak devices, which feature simpler geometries.}\par

By considering monotonic radial profiles of the safety factor and the plasma parallel velocity, along with an electric field composed of a radial equilibrium part with a non-monotonic profile, the twist condition of the dynamical system was violated. The plasma equilibria were perturbed by the superposition of electrostatic harmonic waves, and shearless transport barriers were observed to inhibit chaotic transport. Partial barriers and KAM tori were also found, contributing to the chaotic transport reduction.\par

We observed that, in general, transport barriers become more resilient to perturbations as the Larmor radius increases. With a large Larmor radius, transport barriers are destroyed only with high perturbation amplitudes. Furthermore, even in the absence of barriers, we found that the Larmor radius effect also reduces chaotic transport by making the orbits typically spend more time to escape and reducing the fraction of escaping orbits.\par    

In particular, we have explored the behavior of transport barriers and chaotic transport by examining bifurcation diagrams of shearless transport barriers and the escape rate of an ensemble of chaotic orbits. We analyzed the influence of the electrostatic perturbation amplitude and the impact of the Larmor radius. Regarding the bifurcation diagrams, while varying the control parameters, we examined the rotation number profiles and identified the existence of extreme values where shearless transport barriers can be detected. For the escape rate, we computed the fraction of orbits able to escape from one region to another and the mean escape time of these orbits. \par 

Our results indicate that for small Larmor radii, bifurcations of the shearless curve are likely to occur at high perturbation values, leading to the observation of multiple shearless barriers. However, as the Larmor radius increases, these bifurcations are mitigated. Additionally, we identified intervals of zero escape rate where no shearless transport barriers were found, meaning that other types of barriers emerge to inhibit chaotic transport, such as KAM tori and strong partial barriers. As the Larmor radius increases, the intervals of the electrostatic perturbation amplitude for which some transport barrier exists become greater. In particular, the hardest-to-break and easiest-to-restore shearless transport barriers were found at the largest Larmor radius examined.\par

We also discussed some chaotic transport diagnoses and showed that although the transmissivity, which measures the probability of a given chaotic orbit escaping, is a good indicator for characterizing transport, equal transmissivity scenarios do not necessarily translate into equal chaotic transport conditions due to differences in the characteristic mean escape time. Nevertheless, by using the escape rate, we were able to distinguish subtle differences between scenarios. Specifically, we demonstrated that the escape rate decreases as the Larmor radius increases. We surveyed parameter spaces involving the electrostatic perturbation amplitude and the Larmor radius, computing both transmissivity and escape rate.\par 

\textcolor{black}{Our model, which employs oversimplified drift wave physics and a simple geometry, has several limitations. While the spectrum of turbulent electrostatic fluctuations is inherently complex, we simplify the model by focusing on a single spatial mode with a finite number of harmonics, neglecting the radial dependence of the fluctuations. Additionally, the chaotic advection approach neglects the self-consistency of real turbulence, which is expected to significantly influence particle transport. In particular, nonlinear field coupling can greatly reduce the particle diffusion coefficient\cite{manfredi1997}. Moreover, the dimensionality of the dynamics is reduced to one, disregarding key turbulent mechanisms such as energy cascades and Arnold diffusion.}

\textcolor{black}{Despite these limitations, we were able to capture key features that are consistent with more realistic approaches.} In particular, we have shown that the Larmor radius effect plays a crucial role in the dynamics of chaotic transport and the formation of transport barriers in tokamaks. As the Larmor radius increases, we observe a reduction in chaotic transport and an increase in the robustness of transport barriers.


\begin{acknowledgments}
 The authors thank the financial support from the S{\~a}o Paulo Research Foundation (FAPESP, Brazil) under grants Nos. 2018/03211-6 and \mbox{2020/01399-8};  the Brazilian Federal Agency CNPq under grants Nos. 304616/2021-4, 403120/2021-7 and \mbox{301019/2019-3}; and the  Comit{\'e} Fran\c{c}ais d'Evaluation de la Coop{\'e}ration Universitaire et Scientifique avec le Br{\'e}sil (CAPES/COFECUB) under grant No. 88887.675569/2022-00.\par

The Centre de Calcul Intensif d'Aix-Marseille is acknowledged for granting access to its high-performance computing resources.
\end{acknowledgments}

\appendix 

\section{Rotation number convergence}\label{appendix:convergence}

\begin{figure}[htb]
    \centering
    \includegraphics[width = 0.47\textwidth]{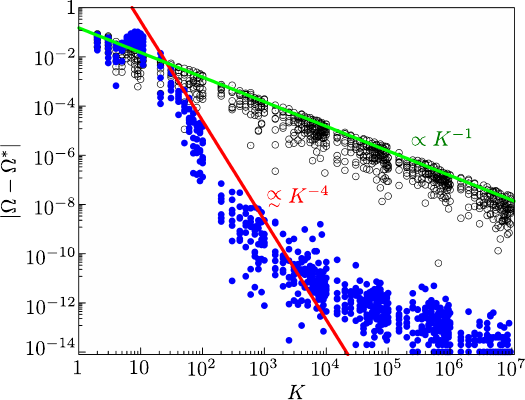}
    \caption{Convergence error of the rotation number, computed using equal-weighted averaging (black empty circles) and the super-convergent method (blue filled circles). The green and red lines show the convergence trends as the integration time increases.}
    \label{fig:error_rotation_number}
\end{figure}

\textcolor{black}{We calculate the convergence error for two numerical methods used to determine the rotation number. Firstly, we compute the rotation number, $\Omega$, by applying an equal-weighted average, using the expressions in \eqref{eq:rotation_number_average} and \eqref{eq:rotation_number_weighter}, with $s(x) = 1$. Then, for comparison, we use the super-convergent method, used in this article, which applies weighted averaging according to \eqref{eq:superconvergent_weighted_average}. In \mbox{Figure \ref{fig:error_rotation_number}}, we compare the convergence errors of both methods, presenting the results for eight orbits near the center of the main resonance of the case shown in \mbox{Figure \ref{fig:phase_space_phi1=0}(a)}, where $I \approx 1$ and the true rotation number is $\Omega^* = 3$.}\par 

\textcolor{black}{As expected, the equal-weighted averaging method shows a convergence proportional to $K^{-1}$, while the super-convergent method achieves a significantly faster convergence, approximately proportional to $K^{-4}$. It is interesting to notice that the super-convergent method exhibits a saturation in convergence, limited by the error tolerance of the numerical integrator, which is set to $10^{-13}$.}

\section{Symplectic map}\label{appendix:map}

 \textcolor{black}{Although in this paper we only discuss the results obtained by the presented model in its ordinary differential form \eqref{eq:model}, we would like to show that an analytical symplectic map can be obtained and that it also represents an interesting topic, mainly, for studying the influence of the FLR effect and the plasma profiles on the chaotic transport of test particles. So, basically, on using the Fourier series representation of the Dirac delta function, $\phi_n = \phi_0$ and \mbox{$\lambda(\psi) = \phi_0\cos(\psi+\alpha)$}, the equations of motion \eqref{eq:model} can be written as}

\textcolor{black}{
\begin{subequations} \label{eq:model_dirac}
\begin{equation}
\dot{I} = -2\pi f(I)\dfrac{\mathrm{d}\lambda(\psi)}{\mathrm{d}\psi}\sum_{n=-\infty}^{+\infty}\delta(\omega_0t-2\pi n), 
\end{equation}
\begin{equation}
\dot{\psi} = g(I)  +  2\pi\dfrac{\mathrm{d} f(I)}{\mathrm{d} I} \lambda(\psi)\sum_{n=-\infty}^{+\infty}\delta(\omega_0t-2\pi n).
\end{equation}
\end{subequations}
}

    \textcolor{black}{Additionally, let us define $I_n=I(t_n^-)$ and $\psi_n=\psi(t_n^-)$, with $t_n^- = nT - \epsilon$, $T = 2\pi/\omega_0$ and $\epsilon \rightarrow 0^+$. Integrating over one jump $(t_n^-,t_{n+1}^-)$, we obtain the discrete model}

\textcolor{black}{
\begin{subequations} \label{eq:map}
\begin{equation}
I_{n+1} = I_n - Tf(I_{n+1})\left.\dfrac{\mathrm{d} \lambda(\psi)}{\mathrm{d} \psi}\right\vert_{\psi_n}, 
\end{equation}
\begin{equation}
\psi_{n+1} = \psi_n+Tg(I_{n+1})  +  \left.T\dfrac{\mathrm{d} f(I)}{\mathrm{d} I}\right\vert_{I_{n+1}} \lambda(\psi_n),
\end{equation}
\end{subequations}
}

    \noindent \textcolor{black}{where the implicit form on $I_{n+1}$ ensures the area-preserving nature of the map. Discrete models are useful because they reproduce the characteristic features of their differential counterparts and reduce the computational cost. For these reasons, we strongly encourage researchers to explore the map described by \eqref{eq:map}, as it promises valuable insights and advancements in the field.}

\section*{Data Availability Statement}
The data that support the findings of this study are available from the corresponding author upon reasonable request.

\bibliography{biblio}

\vspace{\fill}

\end{document}